\documentclass[12pt,thmsa]{article}
\usepackage{sw20lart}

\input{tcilatex}
\begin{document}

\author{Fayyazuddin \\
National\ Centre for Physics and Department of Physics\\
Quaid-i-Azam University\\
Islamabad, Pakistan.}
\title{$B\rightarrow \rho \pi ${\LARGE \ decays and final state phases}}
\maketitle

\begin{abstract}
Using the isospin analysis, the fact that penguin is pure $\Delta I=1/2$
transition, the unitarity for tree graph and $C$-invariance of strong
interactions, it is shown that $\delta _{t}=0=\tilde{\delta}_{t},$ $%
r_{-0}=tr_{0-},\delta _{-0}-\delta _{0-}=\pm \pi ,$ $2tr_{f}r_{\bar{f}}\cos
(\delta _{f}-\delta _{\bar{f}})=(r_{f}^{2}+t^{2}r_{\bar{f}}^{2})-r_{-0}^{2},$
where $\delta $'$s$ are final state phases and $r$'$s$ are penguin to tree
ratios defined in the text.

Using the factorization for tree graph as input and the experimental data,
we have obtained the following bounds on $r_{f},r_{\bar{f}},\delta _{f}$ and 
$\delta _{\bar{f}}:0.11\leq r_{f}\leq 0.21,0.18\leq r_{\bar{f}}\leq 0.30;$ $%
11^{\circ }\leq \delta _{f}\leq 57^{\circ },23^{\circ }\leq \delta _{\bar{f}%
}\leq 90^{\circ }$ for the case $z_{f,\bar{f}}=\cos \alpha \cos \delta _{f,%
\bar{f}}<0$. For $z_{f}<0$ and $z_{\bar{f}}>0,$ we obtain the following
bounds for $r_{\bar{f}}$ and $\delta _{\bar{f}}:0.14\leq r_{\bar{f}}\leq
0.46;$ $90^{\circ }\leq \delta _{\bar{f}}\leq 170^{\circ }$. From
experimental data, for the decays $B^{-}\rightarrow \rho ^{-}\pi ^{0}(\rho
^{0}\pi ^{-})$ we get $\epsilon _{-0}=0.28\pm 0.10,\epsilon _{0-}=0.51\pm
0.10,$ $\frac{A_{CP}^{-0}}{A_{CP}^{0-}}=-0.8\pm 0.1$ i.e. $A_{CP}^{-0}$ and $%
A_{CP}^{0-}$ have opposite sign, where $1+\epsilon _{0-},_{0-}=\frac{\left|
T^{-0,0-}+C^{0-,-0}\right| }{\left| T^{-0,0-}\right| }.$ In the naive quark
model the above values imply $a_{2}/a_{1}=0.39\pm 0.14$ and $%
a_{2}/a_{1}=0.37\pm 0.07$ consistent with each other.
\end{abstract}

\section{Introduction}

The decays of $B$ into two light mesons which belong to an octet or a nonet
representation of $SU(3)$ have been extensively investigated experimentally
and analysed theoretically to obtain information about $CP$-violating phases 
$\alpha ,\beta ,\gamma $ and to test theoretical models \cite{Hassan} . As
is well known, $CP$ asymmetries involve not only weak phases but also strong
phases. It is not easy to reliably estimate the final state strong phases.

In the conventional frame work, the decays are analysed in terms of the tree
amplitude ($T$); the color-supressed amplitude ($C$) and the penguin
amplitude ($P$) (loop supressed). The supression factor is expressed in
terms of Wilson coefficents in the effective Lagrangian for $B$ decays viz $%
\frac{\left| C\right| }{\left| T\right| }\sim \frac{a_{2}}{a_{1}}\approx
0.18 $ , $\frac{\left| P\right| }{\left| T\right| }\approx \left| \frac{a_{4}%
}{a_{1}}\right| \left| \frac{V_{tb}V_{td}^{*}}{V_{ub}V_{ud}^{*}}\right|
\approx 0.03\left| \frac{V_{tb}V_{td}^{*}}{V_{ub}V_{ud}^{*}}\right| .$

However for $B\rightarrow \pi \pi $ decays, to fit the data, rather large
values of $\left| \frac{C}{T}\right| $ and $\left| \frac{P}{T}\right| $ are
required; against the spirit of the model. The pions are identical bosons
and their wave functions in the final state must be symmetric. This is
explict in the isospin analysis of these decays. It is not clear how Bose
statistics is satisfied in terms of $T,C$ and $P$ amplitudes. However $\rho $
and $\pi $ being not identical bosons, there is no problem with Bose
statistics in terms of analysing these decays in terms of amplitudes $T,C$
and $P.$ The $B\rightarrow \rho \pi $ decays have been extensively studied 
\cite{2,3}. In particular, the decays $\bar{B}^{0}\rightarrow \rho ^{-}\pi
^{+}(f),$ $\bar{B}^{0}\rightarrow \rho ^{+}\pi ^{-}(\bar{f})$ and $%
B^{-}\rightarrow \rho ^{-}\pi ^{0}(\rho ^{-}\pi ^{0})$ are analysed in terms
of one weak phase $\alpha ,$ six strong phases $\delta _{f},\delta _{\bar{f}%
} $ ,$\delta _{t},\delta _{-0},\delta _{0-}$ and $\tilde{\delta}_{t}$ and
seven parameters: 
\begin{eqnarray}
\frac{T^{\bar{f}}}{T^{f}} &=&\frac{\left| T^{\bar{f}}\right| }{\left|
T^{f}\right| }e^{i\left( \delta _{\bar{f}}^{T}-\delta _{f}^{T}\right)
}\equiv te^{i\delta _{t}}  \nonumber \\
\frac{P^{\bar{f}}}{T^{\bar{f}}} &=&\frac{\left| P^{\bar{f}}\right| }{\left|
T^{\bar{f}}\right| }e^{i\left( \delta _{\bar{f}}^{P}-\delta _{\bar{f}%
}^{T}\right) }\equiv r_{\bar{f}}e^{i\delta _{\bar{f}}}  \nonumber \\
\frac{P^{f}}{T^{f}} &=&\frac{\left| P^{f}\right| }{\left| T^{f}\right| }%
e^{i\left( \delta _{f}^{P}-\delta _{f}^{T}\right) }\equiv r_{f}e^{i\delta
_{f}}  \nonumber \\
\frac{\tilde{T}^{0-}}{\tilde{T}^{-0}} &=&\frac{T^{0-}+C^{0-}}{T^{-0}+C^{-0}}=%
\frac{\left| \tilde{T}^{0-}\right| }{\left| \tilde{T}^{-0}\right| }%
e^{i\left( \delta _{0-}^{\tilde{T}}-\delta _{-0}^{\tilde{T}}\right) }=\left( 
\frac{1+\epsilon _{0-}}{1+\epsilon _{-0}}\right) te^{i\tilde{\delta}_{t}} 
\nonumber \\
\frac{P^{0-}}{\tilde{T}^{0-}} &=&\tilde{r}_{0-}e^{i\left( \delta
_{0-}^{P}-\delta _{0-}^{\tilde{T}}\right) }=\frac{r_{0-}}{1+\epsilon _{0-}}%
e^{i\delta _{0-}}  \nonumber \\
\frac{P^{-0}}{\tilde{T}^{-0}} &=&\tilde{r}_{-0}e^{i\left( \delta
_{-0}^{P}-\delta _{-0}^{\tilde{T}}\right) }=\frac{r_{-0}}{1+\epsilon _{-0}}%
e^{i\delta _{-0}}  \label{1}
\end{eqnarray}
where 
\[
1+\epsilon _{0-}=\frac{\left| \tilde{T}^{0-}\right| }{\left| T^{0-}\right| }%
,1+\epsilon _{-0}=\frac{\left| \tilde{T}^{-0}\right| }{\left| T^{-0}\right| }%
,\sqrt{2}\left| T^{0-}\right| =\left| T^{\bar{f}}\right| ,\sqrt{2}\left|
T^{-0}\right| =\left| T^{f}\right| 
\]
Thus there are thirteen parameters besides the weak phase $\alpha $. There
are seven independent observables viz the $CP$ violating asymmetries $%
A_{CP}^{\pm }$ and $A_{CP}^{\mp }$, the direct $CP$ violation parameter $C$
and dilution parameter $\Delta C,$ the mixing induced $CP$ violation
parameter $S$ and dilution parameter $\Delta S,$ and the total decay rate $%
\Gamma _{\rho ^{\pm }\pi ^{\mp }}$ for $\bar{B}^{0}$ decays and four
observables for $B^{-}$ decays viz two decay rates and two $CP$ asymmetries.

Hence it is required to reduce number of parameters by some theoretical
input. In particular in this paper (see section 2); using the isospin
analysis and the fact that penguin is pure $\Delta I=1/2$ transition, we
have obtained the following relations 
\begin{eqnarray*}
r_{-0} &=&tr_{0-};\delta _{0-}^{P}-\delta _{-0}^{P}=\pm \pi \\
1-\cos (\delta _{f}^{P}-\delta _{\bar{f}}^{P}) &=&\frac{%
r_{-0}^{2}-(r_{f}-tr_{\bar{f}})^{2}}{2tr_{f}r_{\bar{f}}} \\
tr_{\bar{f}}\sin (\delta _{f}^{P}-\delta _{\bar{f}}^{P}) &=&r_{-0}\sin
(\delta _{-0}^{P}-\delta _{f}^{P})
\end{eqnarray*}
Further, using the unitarity for the tree amplitude and $C$-invariance of
strong interactions, it is shown that one possible solution gives $\delta
_{t}=0=\tilde{\delta}_{t}.$ Selecting this solution we have 
\begin{eqnarray*}
\delta _{f}^{P}-\delta _{\bar{f}}^{P} &=&\delta _{f}-\delta _{\bar{f}} \\
\delta _{-0}^{P}-\delta _{0-}^{P} &=&\delta _{-0}-\delta _{0-}=\pm \pi
\end{eqnarray*}
These results are used to analyse the experimental data in section 3.

\section{Isospin Constraints and Final State Phases}

The decay amplitudes in terms of $T,P$ and $C$ for $\bar{B}^{0}$ $%
\rightarrow f,\bar{f}$ decays are given by 
\begin{eqnarray}
\bar{S}_{f} &=&e^{-i\beta }\langle f\left| H\right| \bar{B}^{0}\rangle
\equiv e^{-i\beta }\bar{A}_{f}=e^{-i\beta }\left[ T^{f}e^{-i\gamma
}+P^{f}e^{i\beta }\right]  \nonumber \\
&=&-T^{f}\left[ e^{i\alpha }-r_{f}e^{i\delta _{f}}\right]  \label{2} \\
\bar{S}_{\bar{f}} &=&e^{-i\beta }\bar{A}_{\bar{f}}=-T^{\bar{f}}\left[
e^{i\alpha }-r_{\bar{f}}e^{i\delta _{\bar{f}}}\right]  \label{3}
\end{eqnarray}
For $B^{0}\rightarrow \bar{f}$ and $B^{0}\rightarrow f$ change $\alpha
\rightarrow -\alpha $ in Eqs.$\left( \text{\ref{2}}\right) $ and $\left( 
\text{\ref{3}}\right) $ respectively.

For $B^{-}\rightarrow \rho ^{-}\pi ^{0}$%
\begin{eqnarray}
\bar{S}_{-0} &=&e^{-i\beta }\bar{A}_{-0}=-\left[ T^{-0}e^{i\alpha
}+C^{-0}e^{i\alpha }-P^{-0}\right]  \nonumber \\
&=&-\left[ \tilde{T}^{-0}e^{i\alpha }-P^{-0}\right] =-\tilde{T}^{-0}\left[
e^{i\alpha }-\tilde{r}_{-0}e^{i\delta _{-0}}\right]  \label{4} \\
\bar{S}_{0-} &=&e^{-i\beta }\bar{A}_{0-}=-\tilde{T}^{0-}\left[ e^{i\alpha }-%
\tilde{r}_{0-}e^{i\delta _{0-}}\right]  \label{5} \\
&&  \nonumber
\end{eqnarray}
For $B^{+}\rightarrow \rho ^{+}\pi ^{0},\rho ^{0}\pi ^{+}$ change $\alpha $
to $-\alpha $ in Eqs.$\left( \text{\ref{4}}\right) $ and $\left( \text{\ref
{5}}\right) $ respectively.

Isospin analysis of these decays gives useful constraints on the decay
amplitudes. The effective weak Lagrangian contains both $\Delta I=\frac{1}{2}
$ and $\Delta I=\frac{3}{2}$ parts. As is well known, the decay amplitudes
can be written in terms of four complex amplitudes corresponding to $I=0,I=2$
symmetric isospin wave functions for $\rho \pi $ states and two
antisymmetric isospin wave functions corresponding to $I=1$. Thus we have 
\begin{eqnarray}
\bar{A}_{f} &=&-\frac{1}{2}A_{1}+\frac{1}{\sqrt{6}}A_{0}+\frac{1}{\sqrt{12}}%
A_{2}-\frac{1}{2}A_{1}^{\prime }  \label{10} \\
\bar{A}_{\bar{f}} &=&\frac{1}{2}A_{1}+\frac{1}{\sqrt{6}}A_{0}+\frac{1}{\sqrt{%
12}}A_{2}+\frac{1}{2}A_{1}^{\prime }  \label{11} \\
\bar{A}_{-0} &=&-\frac{1}{\sqrt{2}}A_{1}+\frac{\sqrt{3}}{2\sqrt{2}}A_{2}-%
\frac{1}{2\sqrt{2}}A_{1}^{\prime }  \label{12} \\
\bar{A}_{0-} &=&\frac{1}{\sqrt{2}}A_{1}+\frac{\sqrt{3}}{2\sqrt{2}}A_{2}+%
\frac{1}{2\sqrt{2}}A_{1}^{\prime }  \label{13} \\
\bar{A}_{00} &=&\frac{1}{\sqrt{6}}A_{0}-\frac{1}{\sqrt{3}}A_{2}  \label{14}
\end{eqnarray}
The amplitudes $A_{0},$ $A_{1}$ and $A_{2},A_{1}^{\prime }$correspond to $%
\Delta I=\frac{1}{2}$ and $\Delta I=\frac{3}{2}$ parts of the effective weak
Lagrangian respectively. From Eqs $\left( \text{\ref{10}-\ref{14}}\right) $,
we get the relations 
\begin{eqnarray}
\left( \bar{A}_{f}+\bar{A}_{\bar{f}}\right) -\sqrt{2}\left( \bar{A}_{-0}+%
\bar{A}_{0-}\right) &=&2\bar{A}_{00}  \label{15} \\
\bar{A}_{-0}+\bar{A}_{0-} &=&\frac{\sqrt{3}}{2}A_{2}  \label{15a}
\end{eqnarray}
and 
\begin{equation}
2\left( \bar{A}_{f}-\bar{A}_{\bar{f}}\right) -\sqrt{2}\left( \bar{A}_{-0}-%
\bar{A}_{0-}\right) =-A_{1}^{\prime }  \label{15b}
\end{equation}
Since both $A_{2}$ and $A_{1}^{\prime }$ are pure $\Delta I=\frac{3}{2}$ and
the penguin $P$ is pure $\Delta I=\frac{1}{2}$, we must have 
\begin{eqnarray}
P^{-0}+P^{0-} &=&0  \label{15c} \\
2(P^{f}-P^{\bar{f}})-\sqrt{2}\left( P^{-0}-P^{0-}\right) &=&0  \label{15d}
\end{eqnarray}
From Eqs.(\ref{15c}) and (\ref{15d}), we get 
\begin{eqnarray}
\delta _{-0}^{P}-\delta _{0-}^{P} &=&\pm \pi  \nonumber \\
r_{-0} &=&r_{0-}\frac{\left| T^{0-}\right| }{\left| T^{-0}\right| }
\label{16} \\
&=&tr_{-0}  \nonumber
\end{eqnarray}
and 
\begin{equation}
1-\cos \left( \delta _{f}^{P}-\delta _{\bar{f}}^{P}\right) =\frac{%
r_{-0}^{2}-\left( r_{f}-tr_{\bar{f}}\right) ^{2}}{2tr_{f}r_{\bar{f}}}
\label{20}
\end{equation}
\begin{equation}
tr_{\bar{f}}\sin \left( \delta _{f}^{P}-\delta _{\bar{f}}^{P}\right)
=r_{-0}\sin \left( \delta _{-0}^{P}-\delta _{f}^{P}\right)  \label{21}
\end{equation}

One can get some information about the final state phases $\delta _{f}^{T}$
and $\delta _{\bar{f}}^{T}$ as follows. The factorization ansatz for tree
graph is on strong footing \cite{4}. This combined with a physical picture 
\cite{5} that in the weak decay of $B$ meson, the $b$ quark is converted
into $b\rightarrow u+q+\bar{q}$ and that for the tree graph the
configuration is such that $q$ and $\bar{q}$ essentially go together into
the color singlet states with the third quark recoiling, there is a
significant probability that the system will hadronize as a two body final
state. The strong phase shifts are generated after hadronization by
rescattering. Thus it is reasonable to use unitarity to get information
about strong phases at least for the tree graph.

Unitarity gives 
\begin{equation}
\func{Im}A_{f}^{i}=\sum_{n}M_{nf}^{*}A_{n}^{i}  \label{25}
\end{equation}
where $A_{f}$ is the decay amplitude and $M_{nf}$ is the scattering
amplitude for $f\rightarrow n.$ The superscript $i$ indicates that the Eq. (%
\ref{25}) holds for each weak amplitude and not for the whole amplitude $%
A_{f}.$ Eq.(\ref{25}) can be used to estimate the final state phase shifts 
\cite{6}. Eq. (\ref{15d}) can be written as 
\begin{equation}
\func{Im}A_{f}^{i}-M_{ff}^{*}A_{f}^{i}=\sum_{n}M_{nf}^{*}A_{n}  \label{26}
\end{equation}
Since the decays are $P$-wave decays, we can use $M=\frac{S-1}{2i}$ where $S$
is the $S$-matrix for $l=1$ partial wave. In terms of $S$, Eq. (\ref{25})
gives 
\begin{equation}
\func{Im}A_{f}^{i}\left( 1+S_{f}^{*}\right) +i\func{Re}A_{f}^{i}\left(
1-S_{f}^{*}\right) =i\sum_{n}S_{nf}^{*}A_{n}^{i}  \label{27}
\end{equation}
Now parametrizing $S$-matrix as $\eta e^{2i\Delta }$ \cite{7} and noting $%
\func{Re}A_{f}^{i}=\left| A_{f}^{i}\right| \cos \delta _{f}^{i},$ $\func{Im}%
A_{f}=\left| A_{f}\right| \sin \delta _{f}$ and taking the absolute square
on both sides, we get 
\begin{equation}
\left| A_{f}^{i}\right| ^{2}\left[ \left( 1+\eta ^{2}\right) -2\eta \cos
2\left( \delta _{f}^{i}-\Delta \right) \right] =\sum_{n^{\prime },n\neq
f}A_{n}^{i}S_{nf}^{*}A_{n^{\prime }}^{i*}S_{n^{\prime }f}  \label{28}
\end{equation}
Note that in single channel description $\eta $ the absorption coefficent
take care of all the inelastic channels \cite{8}. This is an exact equation.
In the random phase approximation of \cite{6}, we can put 
\begin{equation}
\sum_{n^{\prime },n\neq f}A_{n}^{i}S_{nf}^{*}A_{n^{\prime
}}^{i*}S_{n^{\prime }f}=\sum_{n\neq f}\left| A_{n}^{i}\right| ^{2}\left|
S_{nf}\right| ^{2}=\overline{\left| A_{n}^{i}\right| ^{2}}\left( 1-\eta
^{2}\right)  \label{29}
\end{equation}
Then using Eq.(\ref{29}), Eq.(\ref{28}) can be written in the form \cite{7} 
\begin{equation}
\tan ^{2}\left( \delta _{f}^{i}-\Delta \right) =\left( \frac{1-\eta }{1+\eta 
}\right) \frac{\rho ^{2}-\frac{1-\eta }{1+\eta }}{1-\rho ^{2}\frac{1-\eta }{%
1+\eta }}  \label{32}
\end{equation}
where 
\begin{equation}
\rho ^{2}=\frac{\overline{\left| A_{n}^{i}\right| ^{2}}}{\left|
A_{f}^{i}\right| ^{2}}  \label{32a}
\end{equation}
\[
\frac{1-\eta }{1+\eta }\leq \rho ^{2}\leq 1 
\]
Using the $C$-invariance of $S$-matrix \cite{9} 
\begin{equation}
\langle f\left| S\right| n\rangle =\langle f\left| C^{-1}CSC^{-1}C\right|
n\rangle =\langle \bar{f}\left| S\right| \bar{n}\rangle  \label{33}
\end{equation}
we get 
\begin{equation}
\tan ^{2}\left( \delta _{\bar{f}}^{i}-\Delta \right) =\left( \frac{1-\eta }{%
1+\eta }\right) \frac{\bar{\rho}^{2}-\frac{1-\eta }{1+\eta }}{1-\bar{\rho}%
^{2}\frac{1-\eta }{1+\eta }}  \label{34}
\end{equation}
where 
\begin{equation}
\bar{\rho}^{2}=\frac{\overline{\left| A_{\bar{n}}^{i}\right| ^{2}}}{\left|
A_{\bar{f}}^{i}\right| ^{2}}  \label{35}
\end{equation}
In particular we use Eqs.(\ref{32}) and (\ref{34}) for the tree amplitudes $%
i=T.$ First we note that for the minimum value of $\rho ^{2}$ and $\bar{\rho}%
^{2},$ Eq.(\ref{32}) and (\ref{34}) gives 
\begin{equation}
\delta _{f,\bar{f}}^{T}=\Delta \text{ or }\pm \pi +\Delta  \label{35a}
\end{equation}
Thus either we have $\delta _{t}\equiv \delta _{\bar{f}}^{T}-\delta
_{f}^{T}=0$ or $\delta _{t}=\pm \pi $. We select the solution $\delta
_{t}=0\,$in agreement with the result of refrence \cite{10}$.$ This
reinforces the point of view that for the tree graph final state phases
shift is generated by the rescattering.

However if we assume $\rho ^{2}=\bar{\rho}^{2},\,$then from Eqs.(\ref{32})
and (\ref{34}), we get 
\begin{equation}
\tan \left( \delta _{f}^{T}-\Delta \right) =\pm \tan \left( \delta _{\bar{f}%
}^{T}-\Delta \right)  \label{36}
\end{equation}
Then from Eq. (\ref{36}), it follows that 
\begin{eqnarray*}
(i)\text{ }\delta _{f}^{T} &=&\delta _{\bar{f}}^{T}\text{; }\delta _{\bar{f}%
}^{T}-\text{ }\delta _{f}^{T}\equiv \delta _{t}=0 \\
\left( ii\right) \text{ }\delta _{\bar{f}}^{T}-\Delta &=&\pi +\left( \delta
_{f}^{T}-\Delta \right) \\
\left( iii\right) \text{ }\delta _{\bar{f}}^{T}-\Delta &=&-\left( \delta
_{f}^{T}-\Delta \right) ,\text{ } \\
\left( iv\right) \text{ }\delta _{\bar{f}}^{T}-\Delta &=&\pm \pi -\left(
\delta _{f}^{T}-\Delta \right) ,\text{ }
\end{eqnarray*}
Thus besides other solutions for which $\delta _{t}$ is arbitrary, $\delta
_{t}=0$ is also a possible solution.

We now discuss the final state phases for the amplitudes $\tilde{T}^{-0}$
and $\tilde{T}^{0-}$. Since the tree and color supressed amplitudes have
same weak phase, we can use similar analysis as above to obtain the solution 
$\delta _{0-}^{\tilde{T}}$ $=\delta _{-0}^{\tilde{T}}$ or $\delta _{0-}^{%
\tilde{T}}=\mp \pi +\tilde{\delta}_{-0}^{T}$. Now using $\delta
_{f}^{T}=\delta _{\bar{f}}^{T}$ and $\delta _{0-}^{\tilde{T}}=\delta _{-0}^{%
\tilde{T}}$ we get from Eqs.(\ref{16}) and (\ref{20}) 
\begin{eqnarray*}
\delta _{-0}-\delta _{0-} &=&\pm \pi \\
1-\cos (\delta _{f}-\delta _{\bar{f}}) &=&\frac{r_{-0}^{2}-(r_{f}-tr_{\bar{f}%
})^{2}}{2tr_{f}r_{\bar{f}}}
\end{eqnarray*}
We note that $\delta _{f}=\delta _{\bar{f}}$ if $r_{-0}=\pm (r_{f}-tr_{\bar{f%
}})$ which is experimentally testable.

\section{Observables}

In this section, we define the observables which are experimentally measured.

1) \textbf{The average decay rate}

Define 
\begin{eqnarray}
R_{f} &=&\frac{\left| S_{\bar{f}}\right| ^{2}+\left| \bar{S}_{f}\right| ^{2}%
}{2}=\frac{1}{2}\left( \Gamma _{\bar{f}}+\bar{\Gamma}_{f}\right)  \nonumber
\\
R_{\bar{f}} &=&\frac{\left| S_{f}\right| ^{2}+\left| \bar{S}_{\bar{f}%
}\right| ^{2}}{2}=\frac{1}{2}\left( \Gamma _{f}+\bar{\Gamma}_{\bar{f}}\right)
\label{40} \\
R_{-0} &=&\frac{\left| S_{+0}\right| ^{2}+\left| \bar{S}_{-0}\right| ^{2}}{2}
\nonumber \\
R_{0-} &=&\frac{\left| S_{0+}\right| ^{2}+\left| \bar{S}_{0-}\right| ^{2}}{2}
\nonumber
\end{eqnarray}
where 
\begin{eqnarray}
R_{f} &=&\left| T^{f}\right| ^{2}\left[ 1-2r_{f}\cos \alpha \cos \delta
_{f}+r_{f}^{2}\right] \equiv \left| T^{f}\right| ^{2}B_{f}  \nonumber \\
R_{\bar{f}} &=&\left| T^{\bar{f}}\right| ^{2}\left[ 1-2r_{\bar{f}}\cos
\alpha \cos \delta _{\bar{f}}+r_{\bar{f}}^{2}\right] \equiv \left| T^{\bar{f}%
}\right| ^{2}B_{\bar{f}}  \label{41} \\
R_{-0} &=&\left| T^{-0}\right| ^{2}\left[ \left( 1+\epsilon _{-0}\right)
^{2}-2\left( 1+\epsilon _{-0}\right) r_{-0}\cos \alpha \cos \delta
_{-0}+r_{-0}^{2}\right] \equiv \left| T^{-0}\right| ^{2}B_{-0}  \nonumber \\
&&  \label{42} \\
R_{0-} &=&\left| T^{-0}\right| ^{2}\left[ \left( 1+\epsilon _{0-}\right)
^{2}-2\left( 1+\epsilon _{0-}\right) r_{0-}\cos \alpha \cos \delta
_{0-}+r_{0-}^{2}\right] \equiv \left| T^{0-}\right| ^{2}B_{0-}  \nonumber
\end{eqnarray}
It is convenient to define, the average rates for $B$-decays to $\rho
^{+}\pi ^{-}$ and $\rho ^{-}\pi ^{+}$%
\begin{eqnarray}
\Gamma ^{\pm } &=&\frac{\left| S_{\bar{f}}\right| ^{2}+\left| \bar{S}_{\bar{f%
}}\right| ^{2}}{2}  \nonumber \\
\Gamma ^{\mp } &=&\frac{\left| S_{f}\right| ^{2}+\left| \bar{S}_{f}\right|
^{2}}{2}  \label{44}
\end{eqnarray}
\begin{eqnarray}
R_{f}+R_{\bar{f}} &=&\Gamma ^{\rho \pi }=\left( \Gamma ^{\pm }+\Gamma ^{\mp
}\right)  \nonumber \\
\left| T^{f}\right| ^{2} &=&\frac{\Gamma ^{\pm }+\Gamma ^{\mp }}{%
B_{f}+t^{2}B_{\bar{f}}}=\frac{R_{f}}{B_{f}}  \label{44a}
\end{eqnarray}

2) $CP$\textbf{-Violating Asymmetries:- }Define direct $CP$ asymmetries 
\begin{eqnarray}
-A_{CP}^{\pm } &=&a_{f}=\frac{\left| S_{\bar{f}}\right| ^{2}-\left| \bar{S}%
_{f}\right| ^{2}}{\left| S_{\bar{f}}\right| ^{2}+\left| \bar{S}_{f}\right| }=%
\frac{2r_{f}\sin \alpha \sin \delta _{f}}{B_{f}}  \nonumber \\
-A_{CP}^{\mp } &=&a_{\bar{f}}=\frac{2r_{\bar{f}}\sin \alpha \sin \delta _{%
\bar{f}}}{B_{\bar{f}}}  \label{45} \\
-A_{CP}^{-0} &=&a_{-0}=\frac{2r_{-0}(1+\epsilon _{-0})\sin \alpha \sin
\delta _{-0}}{B_{-0}}  \nonumber \\
-A_{CP}^{0-} &=&a_{0-}=\frac{2r_{0-}(1+\epsilon _{0-})\sin \alpha \sin
\delta _{0-}}{B_{0-}}  \label{45a}
\end{eqnarray}
Thus it follows from Eqs.(\ref{41})$,$(\ref{44}) and (\ref{45}) 
\begin{eqnarray}
\Gamma ^{\pm } &=&\left( R_{f}+R_{\bar{f}}\right) \left( 1+A_{CP}\right) 
\nonumber \\
\Gamma ^{\mp } &=&\left( R_{f}+R_{\bar{f}}\right) \left( 1-A_{CP}\right)
\label{46}
\end{eqnarray}
\begin{eqnarray}
A_{CP} &=&\frac{R_{f}a_{f}-R_{\bar{f}}a_{\bar{f}}}{R_{f}+R_{\bar{f}}}=\frac{%
\Gamma ^{\pm }-\Gamma ^{\mp }}{\Gamma ^{\pm }+\Gamma ^{\mp }}  \nonumber \\
&=&\frac{2\sin \alpha \left[ r_{f}\sin \delta _{f}-t^{2}r_{\bar{f}}\sin
\delta _{\bar{f}}\right] }{B_{f}+t^{2}B_{\bar{f}}}  \label{47}
\end{eqnarray}
In order to discuss the mixing induced $CP$ asymmetries we first give a
general expression for the time dependent decay rates for $B^{0}\rightarrow 
\bar{f},f$ in terms of these asymmetries 
\begin{equation}
\Gamma \left( B^{0}\left( t\right) \rightarrow \bar{f},f\right) =\left. 
\frac{e^{-\Gamma t}}{2}\left( R_{f}+R_{\bar{f}}\right) \left( 1\pm
A_{CP}\right) \left[ 
\begin{array}{c}
1+\left( C\pm \Delta C\right) \cos \Delta mt \\ 
-\left( S\pm \Delta S\right) \sin \Delta mt
\end{array}
\right] \right.  \label{48}
\end{equation}
where 
\begin{equation}
C\pm \Delta C=\frac{\left| S_{\bar{f},f}\right| ^{2}-\left| \bar{S}_{\bar{f}%
,f}\right| ^{2}}{\left| S_{\bar{f},f}\right| ^{2}+\left| \bar{S}_{\bar{f}%
,f}\right| ^{2}}=\frac{\pm (R_{f}-R_{\bar{f}})+(R_{f}a_{f}+R_{\bar{f}}a_{%
\bar{f}})}{(R_{f}+R_{\bar{f}})(1\pm A_{CP})}  \label{49}
\end{equation}
The decay rates $\bar{\Gamma}_{\bar{f},f}$ can be obtained from Eq.(\ref{48}%
) by changing $\cos \Delta mt\rightarrow \cos \Delta mt,\sin \Delta
mt\rightarrow -\sin \Delta mt$

From Eqs.(\ref{45}) ,(\ref{47})and (\ref{49})$,$ we obtain 
\begin{eqnarray}
R_{f} &=&\frac{1}{2}\left( R_{f}+R_{\bar{f}}\right) \left[ \left( 1+\Delta
C\right) +CA_{CP}\right]  \nonumber \\
R_{\bar{f}} &=&\frac{1}{2}\left( R_{f}+R_{\bar{f}}\right) \left[ \left(
1-\Delta C\right) -CA_{CP}\right]  \label{54} \\
A_{CP}^{\pm } &=&-a_{f}=-\frac{C+A_{CP}\left( 1+\Delta C\right) }{\left(
1+\Delta C\right) +CA_{CP}}  \label{55} \\
A_{CP}^{\mp } &=&-a_{\bar{f}}=-\frac{C-A_{CP}\left( 1-\Delta C\right) }{%
\left( 1-\Delta C\right) -CA_{CP}}  \label{56} \\
&&  \nonumber
\end{eqnarray}
The mixing induced $CP$ asymmetry $S$ and $\Delta S$ are given by 
\begin{eqnarray}
\left( R_{f}+R_{\bar{f}}\right) \left( 1+A_{CP\;}\right) \left( S+\Delta
S\right) &=&2\func{Im}S_{\bar{f}}^{*}\bar{S}_{\bar{f}}  \nonumber \\
&=&\frac{2t(R_{f}+R_{\bar{f}})}{B_{f}+tB_{\bar{f}}}\left[ 
\begin{array}{c}
\sin \left( 2\alpha +\delta _{t}\right) -r_{\bar{f}}\sin \left( \alpha
+\delta _{\bar{f}}+\delta _{t}\right) \\ 
-r_{f}\sin \left( \alpha -\delta _{f}+\delta _{t}\right) \\ 
+r_{f}r_{\bar{f}}\sin \left( \delta _{\bar{f}}-\delta _{f}+\delta _{t}\right)
\end{array}
\right]  \nonumber \\
&&  \label{58}
\end{eqnarray}
\begin{eqnarray}
\left( R_{f}+R_{\bar{f}}\right) \left( 1-A_{CP\;}\right) \left( S-\Delta
S\right) &=&2\func{Im}S_{f}^{*}\bar{S}_{f}  \nonumber \\
&=&\frac{2t(R_{f}+R_{\bar{f}})}{B_{f}+tB_{\bar{f}}}\left[ 
\begin{array}{c}
\sin \left( 2\alpha -\delta _{t}\right) -r_{\bar{f}}\sin \left( \alpha
-\delta _{\bar{f}}-\delta _{t}\right) \\ 
-r_{f}\sin \left( \alpha +\delta _{f}-\delta _{t}\right) \\ 
+r_{f}r_{\bar{f}}\sin \left( \delta _{f}-\delta _{\bar{f}}-\delta _{t}\right)
\end{array}
\right]  \nonumber \\
&&  \label{59}
\end{eqnarray}

\section{$CP$- Asymmetries and Bound on $r_{f},r_{\bar{f}}$}

In this section, we try to extract information about $r_{f}$ and $r_{\bar{f}%
} $ without assuming $\delta _{f}=\delta _{\bar{f}}$. From reference \cite
{11}, we have 
\begin{eqnarray}
C &=&0.30\pm 0.13\text{ , }\Delta C=0.33\pm 0.13  \nonumber \\
a_{f} &=&0.15\pm 0.08  \label{4.1} \\
a_{\bar{f}} &=&0.53\pm 0.30  \nonumber \\
\Gamma ^{\rho \pi } &=&R_{f}+R_{\bar{f}}=(22.8\pm 2.5)\times 10^{-6}
\label{4.1a}
\end{eqnarray}
From the above experimental values, using Eqs. (\ref{54}-\ref{56}), we
obtain 
\begin{eqnarray}
A_{CP} &=&-0.087\pm 0.07  \nonumber \\
R_{f} &=&(14.9\pm 2.2)\times 10^{-6}  \label{4.1b} \\
R_{\bar{f}} &=&(7.9\pm 1.8)\times 10^{-6}  \label{4.1c}
\end{eqnarray}
It is intresting to see that using the above values for $R_{f}$ and $R_{\bar{%
f}}:$%
\begin{equation}
\frac{R_{f}-R_{\bar{f}}}{R_{f}+R_{\bar{f}}}=0.31\pm 0.14\approx \Delta C
\label{4.2}
\end{equation}
Since both $r_{f}$ and $r_{\bar{f}}$ are small (of the order of $0.2$), so
as a first approximation, neglecting the terms $O(r_{f\bar{f}}^{2})$; we get
from Eqs.(\ref{45}),(\ref{47}) and (\ref{49}) 
\begin{eqnarray}
A_{CP} &\approx &\frac{2\sin \alpha (r_{f}\sin \delta _{f}-t^{2}r_{\bar{f}%
}\sin \delta _{\bar{f}})}{1+t^{2}}=\frac{a_{f}-t^{2}a_{\bar{f}}}{1+t^{2}}
\label{4.3} \\
C &\approx &\frac{4t^{2}}{(1+t^{2})^{2}}\sin \alpha \left[ r_{f}\sin \delta
_{f}+r_{\bar{f}}\sin \delta _{\bar{f}}\right] =\frac{2t^{2}}{(1+t^{2})^{2}}%
(a_{f}+a_{\bar{f}})  \label{4.4} \\
\Delta C &\approx &\frac{1-t^{2}}{1+t^{2}}-\frac{4t^{2}\cos \alpha }{%
(1+t^{2})^{2}}(r_{f}\cos \delta _{f}-r_{\bar{f}}\cos \delta _{\bar{f}})
\label{4.5}
\end{eqnarray}
If we take $t^{2}=0.52$ then using the experimental values for $a_{f}$ and $%
a_{\bar{f}},$ we get 
\begin{eqnarray}
A_{CP} &=&-0.083\pm 0.07  \nonumber \\
C &=&0.31\pm 0.16  \label{4.6}
\end{eqnarray}
remarkably consistent with the experimental values given above. We also note
that for $t^{2}=0.52,\frac{1-t^{2}}{1+t^{2}}\approx 0.32$ very near to the
experimental value of $\Delta C.$ Eq.(\ref{4.5}), indicates that the second
term on the right hand is very small. The parameter $t$ relates the $%
B\rightarrow \rho $ form factor $A_{0}$ with $B\rightarrow \pi $ form factor 
$f_{+}$ if we assume factorization for the tree graph. Factorization gives 
\begin{eqnarray}
t &=&\frac{f_{\pi }A_{0}(m_{\pi }^{2})}{f_{\rho }f_{+}(m_{\rho }^{2})}
\label{4.6a} \\
T &=&\frac{G_{F}}{\sqrt{2}}\left| V_{ub}\right| \left| V_{ud}\right|
a_{1}f_{\rho }2m_{B}\left| \vec{p}\right| f_{+}(m_{\rho }^{2})  \label{4.6b}
\end{eqnarray}
Now $f_{\rho }=208MeV,f_{\pi }=131MeV$ and the recent value \cite{11} of $%
f_{+}(0)=0.27\pm 0.04.$ Using the above values and $t=0.72,$ we obtain $%
A_{0}(m_{\pi }^{2})=0.31\pm 0.05$.\cite{10,12,13} With $\left| V_{ub}\right|
=(3.35\pm 0.40)\times 10^{-3},f_{+}(m_{\rho }^{2})=0.27\pm 0.04$ and $%
t^{2}=0.52,$ we obtain 
\begin{eqnarray}
\Gamma _{tree}^{f} &=&(15.2\pm 2.9)\times 10^{-6}  \nonumber \\
\Gamma _{tree}^{\bar{f}} &=&(7.9\pm 1.8)\times 10^{-6}  \label{4.6c}
\end{eqnarray}
Using above values of $\Gamma _{tree}^{f}$ and $\Gamma _{tree}^{\bar{f}},$
we get 
\begin{eqnarray}
B_{f}+t^{2}B_{\bar{f}} &=&\frac{\Gamma _{\rho \pi }}{\Gamma _{tree}^{f}}%
=1.50\pm 0.32  \nonumber \\
B_{f} &=&\frac{R_{f}}{\Gamma _{tree}^{f}}=0.98\pm 0.24  \nonumber \\
B_{\bar{f}} &=&\frac{R_{\bar{f}}}{\Gamma _{tree}^{\bar{f}}}=1.00\pm 0.25
\label{4.7}
\end{eqnarray}

However, we note that 
\begin{equation}
r_{f,\bar{f}}=z_{f,\bar{f}}\pm \sqrt{B_{f,\bar{f}}-\left( 1-z_{f,\bar{f}%
}^{2}\right) }  \label{4.18}
\end{equation}

Thus 
\begin{equation}
B_{f,\bar{f}}\geq 1-z_{f,\bar{f}}^{2}  \label{4.19}
\end{equation}
If we take weak phase $\alpha $ \cite{Hassan} in the range $112^{\circ }\geq
\alpha \geq 90^{\circ }$, then it follows from Eq.(\ref{4.19}), that $B_{f,%
\bar{f}}\geq 0.86$.

Due to large uncertainity in the experimental data, it is convenient to
select \ particular value for $f_{+}(m_{\rho }^{2})$ $V_{ub}$ in calculating 
$\Gamma _{tree}^{f}$ consistent with the above constraints. With $%
f_{+}(m_{\rho }^{2})$ $V_{ub}=(0.26)(3.40)\times 10^{-3},$ we get $\Gamma
_{tree}^{f}=14.5\times 10^{-6}.$ With this value for $\Gamma _{tree}^{f},$
we obtain 
\begin{equation}
B_{f}=1.03\pm 0.12,\text{ }B_{\bar{f}}=1.05\pm 0.12  \label{4.20}
\end{equation}
However if $z_{f,\bar{f}}<0$ $(0<\delta _{f,\bar{f}}<90^{\circ })$ then it
follows from Eq.(\ref{4.18}) $Bf,\bar{f}>1.$ Since largest errors are in $%
a_{f}$ and $a_{\bar{f}},$ it is reasonable to take some fixed values of $%
B_{f},B_{\bar{f}}$ within the range given above.

Case(i) $z_{f,\bar{f}}<0,$ $B_{f}=1.07,$ $B_{\bar{f}}=1.13;-0.25\leq z_{f,%
\bar{f}}\leq -0.06$

Then we obtain from Eq. (\ref{4.18}) 
\begin{eqnarray}
0.11 &\leq &r_{f}\leq 0.21  \nonumber \\
0.18 &\leq &r_{\bar{f}}\leq 0.30  \label{4.20a}
\end{eqnarray}
Case(ii) $z_{f}<0,z_{\bar{f}}>0$

For this case, we take $B_{\bar{f}}=0.98,0.14\leq z_{\bar{f}}\leq 0.25.$
With these values we obtain the bounds

\begin{equation}
0.14\leq r_{\bar{f}}\leq 0.46  \label{4.21}
\end{equation}
From Eq.(\ref{45}), we have 
\begin{eqnarray}
\sin \delta _{f} &=&\frac{B_{f}a_{f}}{2r_{f}\sin \alpha }  \nonumber \\
&&  \nonumber \\
\sin \delta _{\bar{f}} &=&\frac{B_{\bar{f}}a_{\bar{f}}}{2r_{\bar{f}}\sin
\alpha }  \label{4.24b}
\end{eqnarray}
Taking $\left\langle \sin \alpha \right\rangle =0.96;$we obtain the
following bounds on $\delta _{f}$ and $\delta _{\bar{f}}$ from Eq(\ref{4.24b}%
) for the case(i) 
\begin{eqnarray}
11^{\circ } &\leq &\delta _{f}\leq 57^{\circ }  \label{4.24c} \\
23^{\circ } &\leq &\delta _{\bar{f}}\leq 90^{\circ }  \nonumber
\end{eqnarray}
Thus in the range $23^{\circ }-57^{\circ },\delta _{f}$ and $\delta _{\bar{f}%
}$ can be equal to each other.

For the case(ii) we get $90^{\circ }<\delta _{\bar{f}}\leq 170^{\circ }$

Hence in the range $11^{\circ }\leq \delta _{f}\leq 57^{\circ }$it is
possible to have a solution$\delta _{\bar{f}}=\pi -\delta _{f}$

For the mixing induced $CP$ asymmetries, we get from Eqs.(\ref{58}) and (\ref
{59})$,$with $\delta _{t}=0$

\begin{eqnarray}
S+A_{CP}\Delta S &=&\frac{t}{B_{f}+t^{2}B_{\bar{f}}}\left[ 2\sin 2\alpha
-2\tan \alpha \left( r_{\bar{f}}z_{\bar{f}}+r_{f}z_{f}\right) \right] 
\nonumber  \label{u1} \\
&=&\frac{t}{B_{f}+t^{2}B_{\bar{f}}}\left[ 2\sin 2\alpha +\tan \alpha \left(
B_{\bar{f}}+B_{f}-2-r_{\bar{f}}^{2}-r_{f}^{2}\right) \right]  \label{4.29} \\
&&  \nonumber \\
&&  \nonumber \\
\Delta S+A_{CP}S &=&\frac{t}{B_{f}+t^{2}B_{\bar{f}}}\left( -\cot \alpha
\right) \left[ 2r_{\bar{f}}\sin \alpha \sin \delta _{\bar{f}}-2r_{f}\sin
\alpha \sin \delta _{\bar{f}}\right]  \nonumber \\
&=&-\cot \alpha \frac{1}{2t}\left[ \left( 1-t^{2}\right) \left(
C+A_{CP}\Delta C\right) -\left( 1+t^{2}\right) A_{CP}\right]  \label{4.30}
\end{eqnarray}
From Eq. (\ref{4.30}) it follows that $\Delta S+A_{CP}S$ is determined by $%
t,C+A_{CP}\Delta C,A_{CP}$ and the weak phase $\alpha $. The parameters $%
t,C,\Delta C$ and $A_{CP}$ are independent of particle mixing and can be
determined from the direct $CP$ violation .Hence it follows that $\Delta
S+A_{CP}S$ depends only one unknown parameter, the weak phase $\alpha $.\
Due to large experimental uncertainities in mixing induced $CP$ asymmetries,
we will not discuss them any further.

Finally for the decays $B^{-}\rightarrow \rho ^{-}\pi ^{0}(\rho ^{0}\pi
^{-}) $, using the experimental values for $R_{-0}$ and $R_{0-}$ , we get 
\begin{eqnarray}
B_{-0} &=&\frac{2R_{-0}}{\Gamma _{tree}^{f}}=1.65\pm 0.26  \nonumber \\
B_{0-} &=&\frac{2R_{0-}}{t^{2}\Gamma _{tree}^{f}}=2.31\pm 0.29  \label{4.30a}
\end{eqnarray}
Now using the results given in Eq.(\ref{16}) we get from Eq.(\ref{45a}) 
\begin{eqnarray}
-A_{CP}^{-0} &=&a_{-0}=-\frac{2tr_{0-}\left( 1+\epsilon _{-0}\right) \sin
\alpha \sin \delta _{0-}}{B_{0-}}  \nonumber \\
&\approx &-\frac{2tr_{0-}\sin \alpha \sin \delta _{0-}}{\left( 1+\epsilon
_{-0}\right) }  \label{4.31} \\
-A_{CP}^{0-} &=&a_{0-}=\frac{2r_{0-}\left( 1+\epsilon _{0-}\right) \sin
\alpha \sin \delta _{0-}}{B_{0-}}  \nonumber \\
&\approx &\frac{2tr_{0-}\sin \alpha \sin \delta _{0-}}{\left( 1+\epsilon
_{0-}\right) }  \label{4.32}
\end{eqnarray}
Hence 
\begin{equation}
\frac{A_{CP}^{-0}}{A_{CP}^{0-}}\approx -t\frac{1+\epsilon _{0-}}{1+\epsilon
_{-0}}  \label{4.33}
\end{equation}
Now using the approximation $B_{-0}\approx \left( 1+\epsilon _{-0}\right)
^{2},$ $B_{0-}\approx \left( 1+\epsilon _{0-}\right) ^{2}$ we obtain from
the Eq.(\ref{4.30a}) 
\begin{eqnarray}
\left( 1+\epsilon _{-0}\right) ^{2} &\approx &1.28\pm 0.10\Rightarrow
\epsilon _{-0}=0.28\pm 0.10  \nonumber \\
\left( 1+\epsilon _{0-}\right) ^{2} &\approx &1.51\pm 0.10\Rightarrow
\epsilon _{0-}=0.51\pm 0.10  \label{4.34} \\
\frac{A_{CP}^{-0}}{A_{CP}^{0-}} &=&-0.8\pm 0.1\text{ }\left[ -\frac{0.15\pm
0.12}{0.07\pm 0.13}=-2\pm 4\right] _{\exp }  \label{4.35}
\end{eqnarray}
The following comments are in order. Naively in the factorization ansatz: $%
\epsilon _{0-}\sim \frac{\left| C^{0-}\right| }{\left| T^{0-}\right| }=\frac{%
1}{t}\frac{a_{2}}{a_{1}};\epsilon _{-0}\sim \frac{\left| C^{-0}\right| }{%
\left| T^{-0}\right| }=t\frac{a_{2}}{a_{1}}.$ Using the above values of $%
\epsilon _{-0}$, $\epsilon _{0-}$ and $t=0.72,$ we get $\frac{a_{2}}{a_{1}}%
=0.39\pm 0.14,$ $\frac{a_{2}}{a_{1}}=0.37\pm 0.07$ consistent with each
other.

\textbf{To Conclude: }Using the unitarity equation for the tree amplitude,
and the fact that penguin is pure $\Delta I=\frac{1}{2}$ transition, we have
derived the following constraints on the final state phase shifts:$\delta
_{t}=0,$ $1-\cos (\delta _{f}-\delta _{\overline{f}})=\frac{%
r_{-0}^{2}-(r_{f}-tr_{\overline{f}})^{2}}{2tr_{f}r_{\bar{f}}},\delta
_{-0}-\delta _{0-}=\pm \pi ,$ $r_{-0}=tr_{0-,}$ $tr_{\bar{f}}\sin (\delta
_{f}^{P}-\delta _{\bar{f}}^{P})=r_{-0}\sin (\delta _{-0}^{P}-\delta _{f})$.
From the experimental data, using the factorization for the tree amplitude
and $\alpha =\left( 99_{-9}^{+13}\right) ^{\circ },$ we get the following
bounds on the phase shifts : $11^{\circ }\leq \delta _{f}\leq 57^{\circ
};23^{\circ }\leq \delta _{\bar{f}}\leq 90^{\circ }$ for $z_{f,\bar{f}}<0$
whereas for $z_{f}<0$ and $z_{\bar{f}}>0,$ we get $90^{\circ }\leq \delta _{%
\bar{f}}\leq 170^{\circ }.$

Finally in the range $23^{\circ }\leq \delta \leq 57^{\circ },\delta _{f}$
and $\delta _{\bar{f}}$ can be equal to each other. For $\delta _{f}=\delta
_{\bar{f}}$ $r_{-0}=\pm (r_{f}-tr_{\bar{f}}).$ Equality of $\delta _{f}$ and 
$\delta _{\bar{f}}$ may be the consequence `of $C-$invariance of strong
interactions as the final states $\rho ^{-}\pi ^{+}$ and $\rho ^{+}\pi ^{-}$
are $C$-conjugate of each other and cannot be distinguished by strong
interactions. For $B^{-}\rightarrow \rho ^{-}\pi ^{0}(\rho ^{0}\pi ^{-}),$
we get in factorization ansatz $\frac{\left| C^{-0}\right| }{\left|
T^{-0}\right| }\sim 0.28\pm 0.10,$ $\frac{\left| C^{0-}\right| }{\left|
T^{0-}\right| }=0.51\pm 0.10$ which implies $\frac{a_{2}}{a_{1}}=0.39\pm
0.14,\frac{a_{2}}{a_{1}}=0.37\pm 0.07$ consistent with each other. Finally
noting that penguin is pure $\Delta I=1/2,$ we have shown that assymetries $%
A_{CP}^{-0}$ and $A_{CP}^{0-}$ have opposite sign.

\textbf{Acknowledgments}

The author acknowledges a research grant provided by the Higher Education
Commission of Pakistan to him as a Distinguished National Professor.

\end{document}